\documentstyle[twocolumn,fleqn,9pt,psfig]{article}

  \advance\oddsidemargin by -1in
  \advance\oddsidemargin by -6mm
  \advance\evensidemargin by -1in
  \advance\evensidemargin by -6mm
  \advance\topmargin  by -1in
  \advance\topmargin  by 3mm
\begin{document}

\twocolumn[

%\vspace*{2.5cm}

\begin{center}
{\large
{\sf To be published in {\em Synthetic Metals}, \\
Proceedings of the International Conference on Synthetic Metals,
Snowbird, UT, July 28 -- August 2, 1996} \\

\bigskip

{\bf E-print cond-mat/9607199, July 27, 1996}}

\bigskip\bigskip

{\Large Temperature Evolution of the Quantum Hall Effect in 
Quasi-One-Dimensional Organic Conductors}

\bigskip

Hsi-Sheng Goan and Victor M.  Yakovenko\\

\bigskip
Department of Physics and Center for Superconductivity Research,
University of Maryland, College Park, MD 20742, USA\\

\smallskip
\end{center}

{\normalsize\bf Abstract}

   \hspace*{4mm} The Hall conductivity in the magnetic-field-induced
spin-density-wave (FISDW) state of the quasi-one-dimensional organic
conductors (TMTSF)$_2$X at a finite temperature is calculated.  The
temperature dependence of the Hall conductivity is found to be the
same as the temperature dependence of the Fr\"ohlich current of a
regular charge/spin-density wave.  Predicted dependence
$\sigma_{xy}(T)$ can be verified experimentally in the (TMTSF)$_2$X
compounds if all components of the resistivity tensor are measured and
the conductivity tensor is reconstructed.


\bigskip\medskip
]

   Organic metals of the (TMTSF)$_2$X family, where TMTSF is
tetramethyltetraselenafulvalene and X represents an inorganic anion,
are highly anisotropic, quasi-one-dimensional crystals that consist of
parallel conducting chains.  The overlap of the electron wave
functions and the electric conductivity are the highest in the
direction of the chains (the {\bf a} direction) and are much smaller
in the {\bf b} direction perpendicular to the chains.  In this paper,
we neglect the coupling between the chains in the third, {\bf c}
direction, which is weaker than in the {\bf b} direction, and study
the properties of a single layer (the {\bf a}-{\bf b} plane), modeling
(TMTSF)$_2$X as a system of the uncoupled two-dimensional layers.

   When a strong magnetic field is applied perpendicular to the {\bf
a}-{\bf b} plane, the magnetic-field-induced spin-density-wave (FISDW)
appears in the system (see Ref.\ \cite{Yamaji90} for a review).  In
the FISDW phase, the Hall conductivity per one layer, $\sigma_{xy}$,
is quantized at zero temperature as
\begin{equation}
\sigma_{xy}=2Ne^2/h,
\label{2Ne2/h}
\end{equation}
where $e$ is the electron charge, $h=2\pi\hbar$ is the Planck
constant, and $N$ is an integer that characterizes the FISDW.
However, at a finite temperature, because electrons are thermally
excited above the FISDW energy gap, the Hall conductivity is not
quantized. In this paper, we calculate temperature dependence of the
Hall conductivity in the FISDW state.

   To model (TMTSF)$_2$X, let us consider a 2D system that consists of
many chains, parallel to the $x$ axis and equally spaced along the
$y$-axis with the distance $b$.\footnote{The $x$ and $y$ axes
correspond to the {\bf a} and {\bf b} axes of (TMTSF)$_2$X.} The
chains are coupled through the electron tunneling of the amplitude
$t_b$.  To calculate the Hall conductivity, suppose that a magnetic
field $H$ is applied along the $z$ axis perpendicular to the $(x,y)$
plane, and an electric field $E_y$ is applied perpendicular to the
chains.  The electron Hamiltonian in the FISDW state is:\footnote{We
pay no attention to the spin indices, because they are not important
for our purposes.}
\begin{equation}
\hat{\cal H}=-\frac{\hbar^2}{2m}\frac{\partial^2}{\partial x^2} +
2\Delta\cos(Q_x x)  
+ 2t_b\cos[k_y b-G(x-v_{E_y}t)],
\label{Ham}
\end{equation}
where $m$ is the electron mass, $Q_x$ and $\Delta$ are the wave vector
and the amplitude of the FISDW potential, $k_y$ is the electron wave
vector across the chains, $t$ is the time,
\begin{equation}
G=ebH/\hbar c
\label{Gx}
\end{equation}
is the wave vector of the magnetic field, 
\begin{equation}
v_{E_y}=cE_y/H
\label{vEy}
\end{equation}
is the drift velocity in the crossed electric and magnetic fields, and
$c$ is the velocity of light.  Hamiltonian (\ref{Ham}) is written in
the mixed representation, where an electron wave function depends on
the coordinate $x$ along the chains and the momentum $k_y$ across the
chains.  For simplicity, we set the FISDW wave vector across the
chains, $Q_y$, to zero, and neglect the next-nearest-neighbor hopping
term $2t_b'\cos(2k_y b)$ in Hamiltonian (\ref{Ham}).  The electric and
magnetic fields are introduced in Hamiltonian (\ref{Ham}) via the
Peierls--Onsager substitution, $k_y\rightarrow k_y-eA_y/c\hbar$, in
the gauge
\begin{equation}
A_x=A_z=0,\quad A_y=Hx-E_yct.
\label{At}
\end{equation}
It follows from Eq.\ (\ref{Ham}) that, in the presence of the magnetic
field, the hopping {\em across} the chains becomes a periodic
potential {\em along} the chains with the wave vector $G$ (\ref{Gx})
proportional to the magnetic field.  We will refer to this periodic
potential as the ``hopping potential''.  Due to the presence of the
electric field $E_y$, the hopping potential moves along the chains
with the velocity $v_{E_y}$ (\ref{vEy}), whereas the FISDW potential
is assumed to be pinned and does not move.

   Let us linearize the longitudinal dispersion in Hamiltonian
(\ref{Ham}) near the Fermi energy and focus on the electrons whose
momenta are close to the Fermi momenta $+k_{\rm F}$ and $-k_{\rm F}$.
Let us count their momenta from $+k_{\rm F}$ and $-k_{\rm F}$ and
denote their wave functions as $u$ and $w$. In this representation, a
complete electron wave function is a spinor $(u,w)$, and the
Hamiltonian is a $2\times2$ matrix, which can be expanded over the
Pauli matrices $\hat{\tau}_1$, $\hat{\tau}_2$, $\hat{\tau}_3$, and the
unity matrix $\hat{1}$ (which we will not write explicitly in the
following formulas). It is well known
\cite{Yamaji90,Montambaux84b,Lebed85} that the FISDW wave vector
depends on the magnetic field in the following manner:
\begin{equation}
Q_x=2k_{\rm F}-NG=2k_{\rm F}-NebH/\hbar c,
\label{Qx}
\end{equation}
where $N$ is an integer that characterizes the FISDW.  Taking into
account Eq.\ (\ref{Qx}), Hamiltonian (\ref{Ham}) can be rewritten in
the spinor representation as
\begin{equation}
\hat{\cal H}=-i\hbar v_{\rm F}\hat{\tau}_3\frac{\partial}{\partial x} +
\Delta \hat{\tau}_1 e^{i\hat{\tau}_3 NG x} 
+ 2t_b\cos[k_yb-G(x-v_{E_y}t)],
\label{Hlin}
\end{equation}
where $v_{\rm F}=k_{\rm F}/m$ is the Fermi velocity. The last term in
Eq.\ (\ref{Hlin}) can be eliminated by chiral transformation of the
electron wave function:\footnote{This kind of transformation was first
introduced in Ref.\ \cite{Lebed84} that started development of the
FISDW theory.}
\begin{equation}
\left( \begin{array}{c}u \\ w\end{array} \right)
\:\rightarrow\;
\exp\left\{i\hat{\tau}_3 \frac{2t_b}{\hbar\omega_c}
\sin[k_yb-G(x-v_{E_y}t)]\right\}
\left( \begin{array}{c}u \\ w\end{array} \right)\!,
\label{->}
\end{equation}
where 
\begin{equation}
\hbar\omega_c=\hbar v_{\rm F}G=ebHv_{\rm F}/c
\label{wc}
\end{equation}
is the characteristic energy of the magnetic field (the cyclotron
frequency).  In representation (\ref{->}), Hamiltonian (\ref{Hlin})
becomes
\begin{eqnarray}
\hat{\cal H}&=&-i\hbar v_{\rm F}\hat{\tau}_3\frac{\partial}{\partial x} +
\Delta \hat{\tau}_1 \exp(i\hat{\tau}_3 NG x)
\nonumber \\
&&\times\exp\left\{i\hat{\tau}_3 \frac{4t_b}{\hbar\omega_c}
\sin[k_yb-G(x-v_{E_y}t)]\right\}.
\label{Hnot}
\end{eqnarray}
Expanding the periodic function in the last term of Eq.\ (\ref{Hnot})
into the Fourier series, we get the following expression:
\begin{eqnarray}
\hat{\cal H}&=&-i\hbar v_{\rm F}\hat{\tau}_3\frac{\partial}{\partial x}
+ \Delta \hat{\tau}_1 e^{i\hat{\tau}_3 [N(k_yb+Gv_{E_y}t)]}
\nonumber \\
&&\times\sum_n a_{n+N}e^{i\hat{\tau}_3 n[k_yb-G(x-v_{E_y}t)]},
\label{Hsum}
\end{eqnarray}
where the coefficients of the expansion, $a_n$, are the Bessel
functions: $a_n=J_n(4t_b/\hbar\omega_c)$.\footnote{General expression
(\ref{Hsum}) is valid even when the FISDW has a nonzero transverse
wave vector and the transverse dispersion law of the electrons is more
complicated, but the expression for the expansion coefficients $a_n$
would be different in that case.}  The last term in Eq.\ (\ref{Hsum})
is the sum of many sinusoidal potentials whose wave vectors are the
integer multiples of the magnetic wave vector $G$. Each of these
periodic potentials mixes the $+k_{\rm F}$ and $-k_{\rm F}$ electrons
and opens an energy gap at the electron wave vector $k_x$ shifted from
$\pm k_{\rm F}$ by an integer multiple of $G/2$.  The distance in
energy between the gaps is equal to $\hbar\omega_c$ (\ref{wc}).

   The term with $n=0$ in the sum in Eq.\ (\ref{Hsum}) does not depend
on $x$ and opens the gap right at the Fermi level.\footnote{Since, by
introducing the $\pm$ electrons, we have already subtracted the wave
vectors $\pm k_{\rm F}$, the actual wave vector that corresponds to
this term is $2k_{\rm F}$.} When the temperature $T$ is much lower
than the distance between the energy gaps $\hbar\omega_c$:
\begin{equation}
T\ll\hbar\omega_c,
\label{T<<wc}
\end{equation}
only the gap at the Fermi level is important, whereas the other gaps
may be neglected.  Condition (\ref{T<<wc}) is always satisfied in the
relevant temperature range $0\leq T\leq T_c$ (where $T_c$ is the FISDW
transition temperature) in the weak coupling theory of the FISDW,
where $T_c\ll\hbar\omega_c$.  Thus, let us omit all the terms in the
sum in Eq.\ (\ref{Hsum}), except the term with $n=0$:
\begin{equation}
\hat{\cal H}=-i\hbar v_{\rm F}\hat{\tau}_3\frac{\partial}{\partial x}
+ \Delta_{\rm eff}\hat{\tau}_1 e^{i\hat{\tau}_3[N(k_yb+Gv_{E_y}t)]},
\label{HaN}
\end{equation}
\begin{equation}
\Delta_{\rm eff}=a_N\Delta.
\label{Deff}
\end{equation}
This is the so-called single-gap approximation \cite{Montambaux86}. It
was shown explicitly in Ref.\ \cite{Yakovenko91a} that omission of the
gaps located deeply below the Fermi energy does not change the value
of the Hall conductivity, at least at zero temperature.

   By the above sequence of manipulations, we have combined the two
periodic potentials in Eq.\ (\ref{Ham}) into the single effective
potential (\ref{HaN}) that opens a gap at the Fermi level. It follows
from Eq.\ (\ref{HaN}) that the phase $\varphi$ of this effective
potential changes in time:
\begin{equation}
\dot{\varphi}=-NGv_{E_y},
\label{phi.}
\end{equation}
which means that the effective potential moves along the
chains. Since, at zero temperature, all electrons are confined under
the energy gap opened by this potential, the motion of the potential
induces the Fr\"ohlich current \cite{Gruner94b} along the chains:
\begin{equation}
j_x=-\frac{e}{\pi b}\dot{\varphi}.
\label{jxphi}
\end{equation}
Substituting Eqs.\ (\ref{phi.}), (\ref{Gx}), and (\ref{vEy}) into Eq.\
(\ref{jxphi}), we find the quantum Hall effect (QHE) in agreement with
Eq.\ (\ref{2Ne2/h}):
\begin{equation}
j_x=\frac{2Ne^2}{h}E_y.
\label{jxEy}
\end{equation}
To avoid confusion, we wish to emphasize that here the
FISDW is assumed to be immobile, unlike in Ref.\ \cite{Yakovenko93c}
where the influence of the FISDW motion on the QHE was studied. The
effective potential (\ref{HaN}) moves, because it is a combination of
the stationary FISDW potential and the moving hopping potential
(\ref{Ham}).

   Eq.\ (\ref{jxphi}) is a good starting point to discuss the
temperature dependence of the QHE. According to the above
consideration, the Hall conductivity is the Fr\"ohlich conductivity of
the effective periodic potential (\ref{HaN}). Thus, the temperature
dependence of the QHE must be the same as the temperature dependence
of the Fr\"ohlich conductivity. The latter issue was studied in the
theory of a regular charge/spin density wave (CDW/SDW)
\cite{Lee79,Maki90}. At a finite temperature $T$, the electric current
carried by the CDW/SDW condensate is reduced with respect to the
zero-temperature value (\ref{jxphi}) by a factor $f(T)$. The same
factor reduces the condensate Hall effect at a finite temperature:
\begin{equation}
\sigma_{xy}(T)=f(T)\,2Ne^2/h,
\label{f(T)2Ne2/h}
\end{equation}
\begin{equation}
f(T)=1-\int_{-\infty}^\infty \frac{dk_x}{\hbar v_{\rm F}}
\left(\frac{\partial E}{\partial k_x}\right)^{\!2}
\left[-\frac{\partial n_{\rm F}(E/k_{\rm B}T)}{\partial E}\right],
\label{f(T)}
\end{equation}
where $E=\sqrt{(\hbar v_{\rm F}k_x)^2+\Delta_{\rm eff}^2}$ is the
electron dispersion law in the FISDW phase, $k_{\rm B}$ is the
Boltzmann constant, and $n_{\rm F}(\epsilon)=(e^\epsilon+1)^{-1}$ is
the Fermi distribution function.  The last term in Eq.\ (\ref{f(T)})
reflects the fact that normal quasiparticles, thermally excited above
the energy gap, equilibrate with the immobile crystal lattice; thus,
only a fraction of all electrons is carried along the chains by the
moving periodic potential, which reduces the Hall/Fr\"ohlich
current. Derivation of Eq.\ (\ref{f(T)}) is given in Appendix.

\begin{figure}[t]
\centerline{\psfig{file=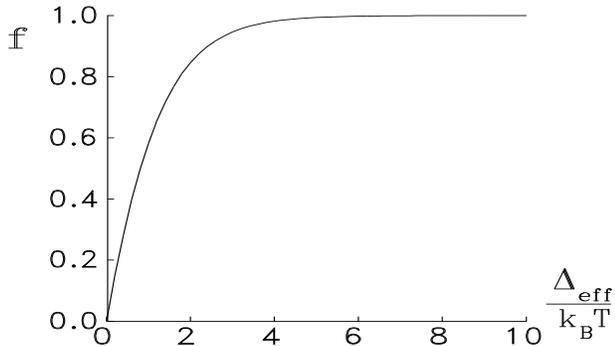,width=\columnwidth,height=0.6\columnwidth,angle=-90}}
\caption{ The reduction factor $f$ of the Hall conductivity as a
function of the ratio of the energy gap at the Fermi level
$\Delta_{\rm eff}$ to the temperature $T$, as given by Eq.\
(\protect\ref{dynMaki}).}
\label{Fig:f(T)}
\end{figure}

\begin{figure}[t]
\centerline {\psfig{file=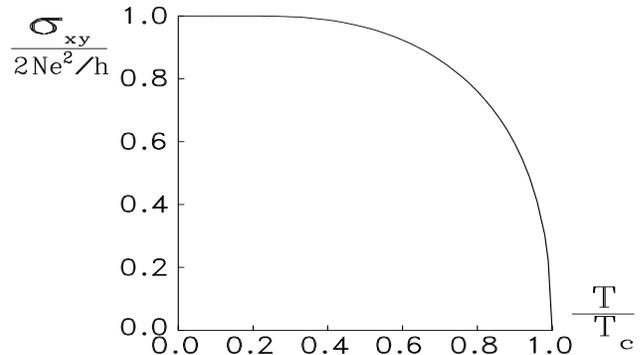,width=\columnwidth,height=0.6\columnwidth,angle=-90}}
\caption{ Hall conductivity in the FISDW state as a function of the
temperature $T$ normalized to the FISDW transition temperature $T_c$.}
\label{Fig:sxy(T)}
\end{figure}

   The function $f$ (\ref{f(T)}) depends only on the ratio of the
energy gap at the Fermi level, $\Delta_{\rm eff}$ (\ref{Deff}), and
the temperature $T$ and can be written as \cite{Maki90,Maki89}
\begin{equation}
f\left(\frac{\Delta_{\rm eff}}{k_{\rm B}T}\right)=
\int_0^\infty d\zeta\,
\tanh\left(\frac{\Delta_{\rm eff}}{2k_{\rm B}T}\cosh\zeta\right)
/\cosh^2\zeta.
\label{dynMaki}
\end{equation}
The function $f(\Delta_{\rm eff}/k_{\rm B}T)$ is plotted in Fig.\
\ref{Fig:f(T)}. It is equal to 1 at zero temperature, where Eq.\
(\ref{f(T)2Ne2/h}) gives the QHE, gradually decreases with increasing
$T$, and vanishes when $T\gg\Delta_{\rm eff}$. Taking into account
that the FISDW order parameter $\Delta$ itself depends on $T$ and
vanishes at the FISDW transition temperature $T_c$, it is clear that
$f(T)$ and $\sigma_{xy}(T)$ vanish at $T\rightarrow T_c$, where
$\sigma_{xy}(T)\propto
f(T)\propto\Delta(T)\propto\sqrt{T_c-T}$. Assuming that the
temperature dependence $\Delta_{\rm eff}(T)$ is given by the BCS
theory \cite{Montambaux86}, we plot the temperature dependence of the
Hall conductivity, $\sigma_{xy}(T)$, in Fig.\ \ref{Fig:sxy(T)}.

   The function $f(T)$ (\ref{f(T)}) is qualitatively similar to the
function $f_{\rm s}(T)$ that describes the temperature reduction of
the superconducting condensate density in the London case. Both
functions approach 1 at zero temperature, but near $T_c$ the
superconducting function behaves differently: $f_{\rm s}(T)\propto
\Delta^2(T)\propto T_c-T$. As explained in Appendix, this is due to
the difference between the static and dynamic limits of the response
function.

   The QHE in the FISDW state at zero temperature was derived
theoretically in Refs.\
\cite{Montambaux84b,Yakovenko91a,Montambaux87}.  An attempt to
calculate the Hall conductivity in the FISDW state at a finite
temperature was made in Ref.\ \cite{Maki89}, but it failed to produce
the QHE at zero temperature. Various aspects of the QHE in
(TMTSF)$_2$X were reviewed in Ref.\ \cite{Yakovenko96}. Temperature
dependence of the Hall resistance in (TMTSF)$_2$X was measured in
experiments \cite{Chaikin92e}. However, to compare the experimental
results with our theory, it is necessary to convert the Hall
resistivity into the Hall conductivity, which requires experimental
knowledge of all components of the resistivity tensor.

   We conclude that, at zero temperature, a FISDW system exhibits the
QHE (\ref{2Ne2/h}) with the same integer number $N$ that characterizes
the wave vector (\ref{Qx}) of the FISDW. As the temperature increases,
the Hall conductivity decreases, vanishing at the FISDW transition
temperature $T_c$. The function $f(T)$ that describes the reduction of
the Hall effect with the temperature is the same as the temperature
reduction function of the Fr\"ohlich current of a regular
charge/spin-density wave.

   This work was partially supported by the NSF under Grant
DMR--9417451, by the Alfred P.~Sloan Foundation, and by the David and
Lucile Packard Foundation.

\section*{Appendix}

   We derived Eqs.\ (\ref{f(T)2Ne2/h}) and (\ref{f(T)}) by calculating
the single-loop Feynman diagram that represents the electromagnetic
response of the electrons in the FISDW state to the electric field
$E_y$. The expression for the diagram is sensitive to the ratio of the
frequency $\omega$ and the wave vector $q$ of the field $E_y$ when
both $\omega$ and $q$ approach zero. Eqs.\ (\ref{f(T)}) and
(\ref{dynMaki}) correspond to the so-called dynamic limit where
$q/\omega=0$ \cite{Maki90}. This limit is appropriate in our case,
because the electric field is, supposedly, strictly homogeneous in
space ($q=0$), but may be time-dependent ($\omega\neq0$). The
effective periodic potential (\ref{HaN}) is also time-dependent.  In
the opposite, static limit $\omega/q=0$, we obtain the function
\begin{equation}
f_{\rm s}(T)=1-\hbar v_{\rm F}\int_{-\infty}^\infty dk_x
\left[-\frac{\partial n_{\rm F}(E/k_{\rm B}T)}{\partial E}\right],
\label{fs(T)}
\end{equation}
which describes the charge-density response to a static deformation of
the CDW phase, $\partial\varphi/\partial x$, as well as the
superconducting condensate density in London superconductors
\cite{Lee79,Maki90}.  In the latter cases, the CDW phase or magnetic
field in the Meissner effect are stationary ($\omega=0$), but vary in
space ($q\neq0$). Comparing Eqs.\ (\ref{f(T)}) and (\ref{fs(T)}), one
can see that $f(T)$ and $f_{\rm s}(T)$ are different. We obtain Eqs.\
(\ref{f(T)}) and (\ref{fs(T)}) by summing over the internal frequency
of the loop first. Different, but equivalent expressions for $f(T)$
and $f_{\rm s}(T)$ were obtained in Ref.\ \cite{Maki90} by integrating
over the internal momentum of the loop first.

   The diagrammatic derivation is not very transparent physically, so
below we offer another derivation of Eq.\ (\ref{f(T)}), based on the
ideas of Refs.\ \cite{Lee79,Overhauser}. Let us consider a
one-dimensional electron system subject to a CDW/SDW of the amplitude
$\Delta_0$, which moves with a small velocity $v_{\rm DW}$. Let us
calculate the Fr\"ohlich current proportional to $v_{\rm DW}$ at a
finite temperature $T$.  We find the electron wave functions in the
reference frame moving with the density wave and then
Galileo-transform them to the laboratory frame \cite{Overhauser}:
\begin{eqnarray}
\psi_k^\pm(x,t)&=&
u_k^{\pm}
e^{i(k_{\rm F}+k+mv_{\rm DW})x-i(k_{\rm F}+k)v_{\rm DW}t\mp iE_kt/\hbar}
\nonumber \\
&&+w_k^{\pm}
e^{i(-k_{\rm F}+k+mv_{\rm DW})x-i(-k_{\rm F}+k)v_{\rm DW}t\mp iE_kt/\hbar},
\label{psi} 
\end{eqnarray}
where we denote $k=k_x$ and keep only the terms linear in $v_{\rm
DW}$. In Eq.\ (\ref{psi}) and below, the index $\pm$ refers to the
states above and below the CDW/SDW energy gap, {\em not} to the states
near $\pm k_{\rm F}$. The coefficients of superposition, $u_k$ and
$w_k$, are given by the following expressions:
\begin{equation}
|u_k^+|^2=|w_k^-|^2=\frac{\Delta_0^2}{2E_k(E_k-\xi_k)},
\label{uw1}
\end{equation}
\begin{equation}
|w_k^+|^2=|u_k^-|^2=\frac{E_k-\xi_k}{2E_k},
\label{uw2} 
\end{equation}
where $\xi_k=\hbar v_{\rm F}k$ and $E_k=\sqrt{\xi_k^2+\Delta_0^2}$ are
the electron dispersion laws in the absence and in the presence of the
CDW/SDW gap.

   By analogy with the standard derivation of the superfluid density
\cite{Landau-IX}, let us assume that, because of interaction with
impurities, phonons, etc., the electron quasiparticles are in thermal
equilibrium with the crystal in the laboratory reference frame, so
their distribution function is the equilibrium Fermi function $n_{\rm
F}$. However, it is not straightforward to apply the Fermi function,
because the two components of the eigenfunction (\ref{psi}), which
have the same energy in the reference frame of the moving CDW/SDW,
have different energies in the laboratory frame. Let us make a
reasonable assumption that a state (\ref{psi}) is populated according
to its {\em average} energy $\bar{E}_k^\pm$:
\begin{eqnarray}
\bar{E}_k^\pm&=&|u_k^\pm|^2(\pm E_k+\hbar(k_{\rm F}+k)v_{\rm DW})
\nonumber \\
&&{}+|w_k^\pm|^2(\pm E_k+\hbar(-k_{\rm F}+k)v_{\rm DW}).
\label{E} 
\end{eqnarray}
The electric current $I$ carried by the electrons is equal to
\begin{eqnarray}
&&I=2e\hbar\sum_{\pm}\int_{-\infty}^\infty\frac{dk}{2\pi}\:
n_{\rm F}\!\left(\frac{\bar{E}_k^\pm}{k_{\rm B}T}\right) \label{I} \\
&& \times\left[|u_k^\pm|^2
\left(\frac{k_{\rm F}+k}{m}+\frac{v_{\rm DW}}{\hbar}\right) +
|w_k^\pm|^2
\left(\frac{-k_{\rm F}+k}{m}+\frac{v_{\rm DW}}{\hbar}\right)\right]\!,
\nonumber
\end{eqnarray}
where the factor 2 comes from the spin. Substituting Eq.\ (\ref{E})
into Eq.\ (\ref{I}) and keeping the terms linear in $v_{\rm DW}$, we
find two contributions to $I$. The first contribution, $I_1$, is
obtained by replacing $\bar{E}_k^\pm$ by $\pm E_k$ in Eq.\ (\ref{I}),
that is, by omitting $v_{\rm DW}$ in Eq.\ (\ref{E}). This term
represents the current produced by all electrons moving with the
velocity $v_{\rm DW}$:
\begin{equation}
I_1=2ev_{\rm DW}2k_{\rm F}/2\pi.
\label{I1}
\end{equation}
The second contribution, $I_2$, comes from expansion of the Fermi
function in Eq.\ (\ref{I}) in $v_{\rm DW}$ and represents reduction of
the current due to thermally excited quasiparticles staying behind the
collective motion:
\begin{eqnarray}
I_2&=&2emv_{\rm DW}\sum_{\pm}\int_{-\infty}^\infty\frac{dk}{2\pi}\:
\frac{\partial n_{\rm F}(\pm E_k/k_{\rm B}T)}{\partial E_k}
\nonumber \\
&& \times\left[v_{\rm F}(|u_k^\pm|^2-|w_k^\pm|^2)
+\frac{\hbar k}{m}(|u_k^\pm|^2+|w_k^\pm|^2)\right]^2.
\label{I2}
\end{eqnarray}
The second term in the brackets in Eq.\ (\ref{I2}) is small compared
to the first term and may be neglected. Substituting Eqs.\ (\ref{uw1})
and (\ref{uw2}) into Eq.\ (\ref{I2}) and expressing the CDW/SDW
velocity in terms of the CDW/SDW phase derivative in time, $v_{\rm
DW}=-\dot{\varphi}/2k_{\rm F}$, we find the temperature-dependent
expression for the Fr\"ohlich current:
\begin{equation}
I=I_1+I_2=-ef(T)\dot{\varphi}/\pi,
\label{Iphi}
\end{equation}
\begin{equation}
f(T)=1-\int_{-\infty}^\infty d\xi_k
\left(\frac{\xi_k}{E_k}\right)^{\!2}
\left[-\frac{\partial n_{\rm F}(E_k/k_{\rm B}T)}{\partial E_k}\right].
\label{ff(T)}
\end{equation}
Eq.\ (\ref{ff(T)}) is the same as Eq.\ (\ref{f(T)}). Dividing the
current per one chain, $I$ (\ref{Iphi}), by the interchain distance
$b$, we get the density of current per unit length, $j_x$
(\ref{jxphi}).


\begin{thebibliography}{99}
\itemsep 0pt

\bibitem{Yamaji90} T. Ishiguro and K. Yamaji, {\em Organic
Superconductors}, Springer-Verlag, Berlin, 1990, Chapter 9.

\bibitem{Montambaux84b} M. H\'eritier, G. Montambaux, and P. Lederer,
{\em J. Phys. Lett. (Paris)}, {\bf 45} (1984) L943.

\bibitem{Lebed85} A.G. Lebed', {\em Zh. Exp. Teor. Fiz.}, {\bf 89}
(1985) 1034 ({\em Sov. Phys. JETP}, {\bf 62} (1985) 595).

\bibitem{Lebed84} L.P. Gor'kov and A.G. Lebed', {\em
J. Phys. Lett. (Paris)}, {\bf 45} (1984) L433.

\bibitem{Montambaux86} D. Poilblanc {\it et~al.}, {\em J. Phys. C},
{\bf 19} (1986) L321; A. Virosztek, L. Chen, and K. Maki, {\em
Phys. Rev. B}, {\bf 34} (1986) 3371.

\bibitem{Yakovenko91a} V.M. Yakovenko, {\it Phys. Rev. B}, {\bf 43}
(1991) 11353.

\bibitem{Gruner94b} G. Gr\"uner, {\it Density Waves in Solids},
Addison-Wesley, New York, 1994.

\bibitem{Yakovenko93c} V.M. Yakovenko, {\it J. Phys. IV (Paris),
Colloque C2}, {\bf 3} (1993) 307; {\it J. Supercond.}, {\bf 7} (1994)
683; V.M. Yakovenko and H.-S. Goan, in {\it Proceedings of the
Physical Phenomena at High Magnetic Fields--II Conference}, Z. Fisk
{\it et~al.} (eds.), World Scientific, Singapore, 1996, p. 116.

\bibitem{Lee79} P.A. Lee and T.M. Rice, {\it Phys. Rev. B}, {\bf 19}
(1979) 3970; T.M. Rice, P.A. Lee, and M.C. Cross, {\it ibid.}, {\bf
20} (1979) 1345.

\bibitem{Maki90} K. Maki and A. Virosztek, {\it Phys. Rev. B}, {\bf
41} (1990) 557; {\bf 42} (1990) 655.

\bibitem{Maki89} A. Virosztek and K. Maki, {\it Phys. Rev. B}, {\bf
39} (1989) 616.

\bibitem{Montambaux87} D. Poilblanc {\it et~al.}, {\it Phys.  Rev.
Lett.}, {\bf 58} (1987) 270; M.Y. Azbel, P. Bak, and P.M. Chaikin,
{\em ibid.}, {\bf 59} (1987) 926.

\bibitem{Yakovenko96} V.M. Yakovenko and H.-S. Goan, cond-mat/9607129,
to be published in {\it J. Phys. I (Paris)}, I.~F.~Schegolev Memorial
Volume.

\bibitem{Chaikin92e} W. Kang {\it et~al.}, {\it Phys. Rev. B}, {\bf
45} (1992) 13566; see also Valfells S. {\it et~al.}, cond-mat/9606212.

\bibitem{Overhauser} M.L. Boriack and A.W. Overhauser, {\it
Phys. Rev. B}, {\bf 15} (1977) 2847; {\bf 16} (1977) 5206; {\bf 17}
(1978) 2395.

\bibitem{Landau-IX} L.D. Landau and E.M. Lifshitz, {\em Statistical
Physics, Part 2}, Pergamon, Oxford, 1991, \S 27.

\end{thebibliography}
\end{document}